# Ion acoustic solitary waves in the four-component complex plasma with a Cairns-Tsallis distribution

Wang Hong, Du Jiulin

*Department of Physics, School of Science, Tianjin University, Tianjin 300350, China*

**Abstract**

We study the ion acoustic solitary waves in the four-component complex plasma consisting of cold inertial ions, positrons, cold and hot (two-temperature) electrons, where the electrons and positrons are the Cairns-Tsallis distribution and have different nonthermal and nonextensive parameters. Base on the plasma hydrodynamic equations and Sagdeev pseudo-potential theory, we derive the conditions of the solitary waves to exist in the plasma, such as the Sagdeev pseudo-potential, the normalized electrostatic potential, the lower and upper limits of Mach number, and the compressive/rarefactive solitary wave. We show that, according to the present study, the solitary wave solutions exist only if the positrons are a Maxwellian distribution and the two-temperature electrons have the same nonextensive and nonthermal parameters. Numerical analyses are made for the conditions of solitary waves depending on the nonextensive and nonthermal parameters in the Cairns-Tsallis distribution.

## 1. Introduction

Ion acoustic wave is a low-frequency longitudinal plasma density oscillation in which electrons and ions propagate in phase space [1,2]. Theoretical and experimental investigations for the dynamic process of ion acoustic waves have been conducted for several decades [3-6]. A large number of studies have revealed that ion acoustic solitary waves and double layers are ubiquitous in various of plasmas, no matter in laboratory plasmas, Earth's magnetosphere, dusty plasmas or in quantum plasmas, where the double layers can accelerate, decelerate or reflect plasma particles [7]. In 2012, Dubinov et al. found a super-soliton wave in very special plasma consisting of five components, and they declared that the wave might also exist in four-component plasmas [8].

The solitary wave theory was developed between the 1960s and 1970s. Theoretical calculations and experimental observations for shallow water waves show that after collisions, the solitary waves retain their original shapes and speeds and the conservation of momentum and energy, which just like the elastic collisions of matter particles. This kind of solitary wave with particle characteristics is called as soliton [1-3]. By a combination of experiments and analytical methods, it has been found that many nonlinear differential equations have the soliton solutions. These results are quickly verified from the fields such as fluid physics, solid-state physics, plasma physics and optical experiments, and they found many practical applications. For example, the low-intensity light pulses used in optical fiber communication are deformed due to dispersion [6], which has low information transmission volume and poor quality, but also adds waveform repeaters at regular intervals on the line. After the "optical soliton" was discovered, these difficulties have been resolved, and the efficiency of information transmission has been improved greatly.

Research on large-scale solitary waves usually employs the Sagdeev pseudo-potential model, which has been widely concerned in the plasma with two-temperature electrons in the early years [9-13]. Two-temperature electron distributions are very common both in the laboratory [11] and in space plasmas [14]. Bharuthram and Shukla inspected the large-amplitude ion acoustic double layers in the unmagnetized three-component plasma with cold ions and two-temperature electrons [12]. Berthomier et al. evaluated the ion acoustic solitary waves and double layers in the unmagnetized three-component plasma with the two-temperature electrons following a Maxwellian distribution [13], and the results were verified by the Viking satellite observation. Baboolal et al. investigated influence of various parameters on the double-layer structure in the plasma with two-temperature electrons and multi-ions [15]. Kaurakis and Shukla studied the enveloping solitary wave in the plasma with two-temperature electrons and cold inertial ions in the magnetosphere [16]. In the study of the ion acoustic double layer with small amplitude, Mishra et al. found that there are two critical concentrations of positrons in the electron-positron-ion plasma with a two-temperature electron distribution [17], which determines the ion acoustic double layer. Verheest and Pillay discussed effects of negatively charged cold dusts and nonthermally distributed ions/electrons on the large-amplitude dust acoustic solitary waves [18], and then they used the hydrodynamic equations to analyze the ion acoustic solitary waves and double layers in the plasma with nonthermal electrons [19].

In recent decades, it has been observed that the velocity distributions of high-energy particles often deviate from the Maxwellian distribution in astrophysical and space plasma environments [20-23]. Spacecraft missions confirmed that existence of excessive high-energy particles can cause enhancement of the high-energy tail [24,25]. Subsequently, different models were developed to describe the non-equilibrium effects in the plasmas. Early in 1968, Vasyliunas introduced the kappa-distribution to simulate the velocity distributions in the space plasmas [26]. In 1995, Cairns et al. introduced a modified velocity distribution function to study the solitary wave structure of nonthermal plasma, [27], where a parameter $\alpha$ ($0 < \alpha < 1$) described the number of super-thermal particles, and the Maxwellian distribution function is recovered only when one takes $\alpha=0$. This model has been applied to study many properties of nonthermal and non-equilibrium space plasmas [4,18,19,28-30].

In recent years, nonextensive statistical theory has been applied to study the nonequilibrium complex plasmas with the power-law velocity distributions [31-42]. And we found that the kappa-distribution observed in space plasmas is actually equivalent to the $q$-distribution in nonextensive statistics as long as we make the parameter translations [41]. A hybrid Cairns-Tsallis distribution was also used to investigate the ion acoustic solitary structures in the complex plasmas [43-45]. Amour et al. studied the electron acoustic soliton structure in the plasma with the nonthermal nonextensive distribution [46]. Williams et al. [47] discussed properties of the plasma media containing excess super-thermal particles. Although multi-component complex plasmas may have power-law velocity distributions, however different component does not have the same nonextensive parameter and the same nonthermal parameter. In this work, based on the Sagdeev pseudo-potential theory, we study the ion acoustic solitary waves in the four-component plasma with ions, positrons and two-temperature electrons which follow the Cairns-Tsallis distributions but have different nonextensive and nonthermal parameters.

In Sec. 2, we give the basic theoretical model and hydrodynamic equations of the plasma and then derive the related physical quantities of the ion acoustic solitary waves. In Sec. 3, we numerically analyze the effects of the nonthermal parameter $\alpha$ and the nonextensive parameter $q$ on the ion acoustic solitary waves. In Sec. 4, we study the small amplitude theory. And in Sec. 5, we give the conclusions. In addition, some complicated mathematical calculations in this paper are given in the Appendix.

## 2. Theoretical Model and Basic Equations

We consider the unmagnetized and nonthermal complex plasma consisting the four components, fluid ions, positrons, cold electrons and hot electrons (i.e. the two-temperature electrons), where the two-temperature electrons and positrons may follow the Cairns-Tsallis distributions in general. We now denote the number densities $n_c$ and $n_h$, the temperatures $T_c$ and $T_h$ of the cold and hot electrons, respectively. The inertia of electrons and positrons in the ion acoustic waves can be ignored. Therefore, the hydrodynamic equations of one-dimensional ion acoustic oscillations can be governed by the normalized dimensionless equations [7,17,48,49],

$$\frac{\partial n_i}{\partial t} + \frac{\partial}{\partial x}(n_i u_i) = 0, \tag{1}$$

$$\frac{\partial u_i}{\partial t} + u_i \frac{\partial u_i}{\partial x} = -\frac{\partial \Psi}{\partial x}, \tag{2}$$

$$\frac{\partial^2 \Psi}{\partial x^2} = n_h + n_c - \eta n_p - (1-\eta) n_i, \tag{3}$$

where, Eq. (1) is the continuity equation, Eq. (2) is the equation of fluid motion, and Eq. (3) is the Poisson equation of the plasma; $u_i$ is the velocity of ions, $n_i$ and $n_p$ are the number density of ions and positrons, respectively; $\Psi = -e\phi/kT_{eff}$ is the normalized electrostatic potential, where $T_{eff}$ is the effective temperature of two-electron components in the plasma, $k$ is Boltzmann constant. If $n_{e0}$, $n_{c0}$, $n_{h0}$, $n_{p0}$ and $n_{i0}$ are the equilibrium density of total electrons, cold electrons, hot electrons, positrons and ions, respectively, and $\eta = n_{p0}/n_{e0}$ is the equilibrium density ratio of positrons to electrons, then in Eqs. (1)-(3) the densities $n_c$, $n_h$ and $n_p$ can be normalized by $n_{e0}$, the ion fluid velocity $u_i$ can be normalized by the effective ion acoustic speed $c_s = (kT_{eff}/m_i)^{1/2}$, and $x$ and $t$ can be normalized by the Debye length $\lambda_{De} = (kT_{eff}/4\pi n_{e0} e^2)^{1/2}$ and the ion plasma period $\varpi_i^{-1} = (4\pi n_{e0} e^2 / m_i)^{-1/2}$, where $m_i$ is mass of the ion and $e$ is charge of electron.

In the four-component complex plasma, positrons and two-temperature electrons are assumed to follow the Cairns-Tsallis distributions but they may have different nonextensive and nonthermal parameters, so they can play different roles in the ion acoustic solitary waves. In the framework of nonextensive statistics based on the probabilistically independent postulate, the entropy and energy are both nonextensive [31]. Therefore when we introduce a potential energy $\varphi(r)$ to the velocity $q$-distribution function, the Tsallis distribution is written [50,51] by

$$f(r,v) \sim \left[1-(q-1)\frac{\varphi(r)}{kT}\right]^{1/(q-1)} \left[1-(q-1)\frac{mv^2}{2kT}\right]^{1/(q-1)}, \tag{4}$$

Combined with the way introducing a potential to the velocity $\alpha$-distribution in the non-thermal plasma by Cairns et al [27], for the present four-component plasma, the one-dimensional Cairns-Tasllis (C-T) distribution of the $j$th component can be written as

$$f(x,v_j) = C_j \left[1 + \alpha_j \left(\frac{m_j v_j^2 - 2Q_j\phi(x)}{kT_j}\right)^2\right] \left[1 - (q_j-1)\frac{m_j v_j^2}{2kT_j}\right]^{\frac{1}{q_j-1}} \left[1 + (q_j-1)\frac{Q_j\phi(x)}{kT_j}\right]^{\frac{1}{q_j-1}}, \quad (5)$$

with a normalization coefficient,

$$C_j = n_{j0}\sqrt{\frac{m_j}{2\pi kT_j}} \begin{cases} \dfrac{(1-q_j)^{5/2}\,\Gamma\left(\dfrac{1}{1-q_j}\right)}{\Gamma\left(\dfrac{1}{1-q_j}-\dfrac{5}{2}\right)\left[3\alpha_j + (1-q_j)^2\left(\dfrac{1}{1-q_j}-\dfrac{3}{2}\right)\left(\dfrac{1}{1-q_j}-\dfrac{5}{2}\right)\right]}, & \dfrac{3}{5} < q_j < 1, \\[2ex] \dfrac{(q_j-1)^{5/2}\left(\dfrac{1}{q_j-1}+\dfrac{5}{2}\right)\left(\dfrac{1}{q_j-1}+\dfrac{3}{2}\right)\Gamma\left(\dfrac{1}{q_j-1}+\dfrac{3}{2}\right)}{\left[(q_j-1)^2\left(\dfrac{1}{q_j-1}+\dfrac{3}{2}\right)\left(\dfrac{1}{q_j-1}+\dfrac{5}{2}\right)+3\alpha_j\right]\Gamma\left(\dfrac{1}{q_j-1}+1\right)}, & q_j > 1, \end{cases}$$

where the subscript $j$ represents the plasma component, i.e., $j = c$, $h$ and $p$ are respectively cold electrons, hot electrons and positrons; $m_j$, $v_j$ and $T_j$ are the mass, velocity and temperature, $Q_j$ is the charge, e.g., $Q_p = e$, $Q_c = Q_h = -e$.

The nonthermal parameter $\alpha_j > 0$ is the nonthermal properties of electrons, e.g., the number of nonthermal electrons in the plasma [27]. The nonextensive parameter $q_j > 0$ can be determined in the nonequilibrium complex plasma by the equation [32],

$$k\frac{dT_j}{dx} + (q_j - 1)Q_j\frac{d\phi}{dx} = 0. \quad (6)$$

Obviously, in the limit of $q_j \to 1$, the C-T distribution (5) becomes the non-thermal α-distribution by Cairns et al. [27], and if $\alpha_j = 0$, it becomes the Tsallis $q$-distribution in nonextensive statistics [50,51]. If we take $q_j \to 1$ and $\alpha_j = 0$, (5) becomes a Maxwellian-Boltzmann distribution. Also for $q_j > 1$, there is a thermal cutoff allowed for the maximum velocity of particles in the function (5), namely,

$$v_{max} = \sqrt{\frac{2kT_j}{m_j(q_j-1)}}. \quad (7)$$

But for $3/5 < q_j < 1$, there is no velocity limit.

The number density of the $j$-th component can be obtained by integrating the distribution function (5) over velocity space [45-47],

$$n_j(\phi) = n_{j0}\left[1 + (q_j-1)\frac{Q_j\phi}{kT_j}\right]^{\frac{1}{q_j-1}}\left[1 + A_j\frac{Q_j\phi}{kT_j} + B_j\left(\frac{Q_j\phi}{kT_j}\right)^2\right], \quad (8)$$

where coefficients $A_j$ and $B_j$ are

$$A_j = -\frac{8\alpha_j(5q_j-3)}{(3q_j-1)(5q_j-3)+12\alpha_j} \quad \text{and} \quad B_j = \frac{4\alpha_j(3q_j-1)(5q_j-3)}{(3q_j-1)(5q_j-3)+12\alpha_j}. \quad (9)$$

When we take $\alpha_j = 0$, (8) becomes the number density in the $q$-distributed plasma [50],

$$n_j(\phi) = n_{j0}\left[1 + (q_j-1)\frac{Q_j\phi}{k_B T_j}\right]^{\frac{1}{q_j-1}}. \quad (10)$$

When we take $q_j \to 1$, (8) becomes the number density in the nonthermal distributed plasma [27],

$$n_j(\phi) = n_{j0}\left[1 + A_j \frac{Q_j\phi}{kT_j} + B_j\left(\frac{Q_j\phi}{kT_j}\right)^2\right]\exp\left(\frac{Q_j\phi}{kT_j}\right). \tag{11}$$

For the nonthermal complex plasma with the two-temperature electrons, if $n_c$, $n_h$ and $T_c$, $T_h$ are the number density and the temperature of the cold electrons and the hot electrons, respectively, $\mu_c = n_{c0}/n_{e0}$ and $\mu_h = n_{h0}/n_{e0}$ are the density ratios of the cold and hot electrons respectively to the total electrons at $\phi = 0$ (so we have $\mu_c + \mu_h = 1$), $\beta = T_c/T_h$ is the temperature ratio of the cold electrons to the hot electrons, and $T_{eff} = T_c/(\mu_c + \mu_h\beta)$ is the effective temperature of cold and hot electrons in the plasma, then using the normalized electrostatic potential $\Psi = -e\phi/kT_{eff}$, we can write the number densities of cold electrons and hot electrons as

$$n_c(\phi) = \mu_c\left[1 - (q_c - 1)\frac{e\phi}{kT_c}\right]^{\frac{1}{q_c-1}}\left[1 - A_c\frac{e\phi}{kT_c} + B_c\left(\frac{Q_c\phi}{kT_c}\right)^2\right]$$
$$= \mu_c\left[1 + \frac{(q_c-1)\Psi}{\mu_c + \mu_h\beta}\right]^{\frac{1}{q_c-1}}\left[1 + \frac{A_c\Psi}{\mu_c + \mu_h\beta} + B_c\left(\frac{\Psi}{\mu_c + \mu_h\beta}\right)^2\right], \tag{12}$$

$$n_h(\phi) = \mu_h\left[1 - (q_h - 1)\frac{e\phi}{kT_h}\right]^{\frac{1}{q_h-1}}\left[1 - A_h\frac{e\phi}{kT_h} + B_h\left(\frac{e\phi}{kT_h}\right)^2\right]$$
$$= \mu_h\left[1 + \frac{(q_h-1)\beta\Psi}{\mu_c + \mu_h\beta}\right]^{\frac{1}{q_h-1}}\left[1 + \frac{A_h\beta\Psi}{\mu_c + \mu_h\beta} + B_h\left(\frac{\beta\Psi}{\mu_c + \mu_h\beta}\right)^2\right], \tag{13}$$

Let $\gamma = T_{eff}/T_p$ be the ratio of the effective temperature to the positron temperature, we write the number density of positrons as

$$n_p(\phi) = \left[1 + (q_p - 1)\frac{e\phi}{k_BT_p}\right]^{\frac{1}{q_p-1}}\left[1 + A_p\frac{e\phi}{k_BT_p} + B_p\left(\frac{e\phi}{k_BT_p}\right)^2\right]$$
$$= \left[1 - (q_p - 1)\gamma\Psi\right]^{\frac{1}{q_p-1}}\left[1 - A_p\gamma\Psi + B_p(\gamma\Psi)^2\right]. \tag{14}$$

In order to study the nonlinear properties of the ion acoustic solitary waves, we consider that all variables $n_i$, $u_i$ and $\Psi$ depend only on a simple variable $\xi = x - Mt$, where $\xi$ is normalized by $\lambda_{De}$ and $M$ is the Mach number ($M$ = the solitary wave speed / $c_s$) [15,29,45,49], and then, Eqs. (1)-(3) can be transformed into the following forms,

$$-M\frac{\partial n_i}{\partial \xi} + \frac{\partial(n_i u_i)}{\partial \xi} = 0, \tag{15}$$

$$-M\frac{\partial u_i}{\partial \xi} + u_i\frac{\partial u_i}{\partial \xi} = -\frac{\partial \Psi}{\partial \xi}, \tag{16}$$

$$\frac{\partial^2 \Psi}{\partial \xi^2} = n_h + n_c - \eta n_p - (1-\eta)n_i. \tag{17}$$

We assume that the perturbation only exists in a finite range. At $|\xi| \to \pm\infty$, the appropriate boundary conditions are given [46] by

$$\Psi \to 0, \quad \frac{d\Psi}{d\xi} \to 0, \quad n_i \to 1, \quad u_i \to 0. \tag{18}$$

Using these boundary conditions to integrate Eq. (15) and Eq.(16), we obtain the number density of ions as

$$n_i = \left(1 - \frac{2\Psi}{M^2}\right)^{-\frac{1}{2}}, \tag{19}$$

where $\Psi < M^2/2$. Substituting Eq. (19) and Eqs. (12)-(14) into Eq. (17), we get that

$$\begin{aligned}\frac{\partial^2 \Psi}{\partial \xi^2} &= \mu_h \left[1 + \frac{(q_h - 1)\beta\Psi}{\mu_c + \mu_h\beta}\right]^{\frac{1}{q_h-1}} \left[1 + A_h \frac{\beta\Psi}{\mu_c + \mu_h\beta} + B_h \left(\frac{\beta\Psi}{\mu_c + \mu_h\beta}\right)^2\right] \\ &+ \mu_c \left[1 + \frac{(q_c - 1)\Psi}{\mu_c + \mu_h\beta}\right]^{\frac{1}{q_c-1}} \left[1 + A_c \frac{\Psi}{\mu_c + \mu_h\beta} + B_c \left(\frac{\Psi}{\mu_c + \mu_h\beta}\right)^2\right] \\ &- \eta \left[1 - (q_p - 1)\gamma\Psi\right]^{\frac{1}{q_p-1}} \left[1 - A_p\gamma\Psi + B_p (\gamma\Psi)^2\right] - \frac{1-\eta}{\sqrt{1 - 2\Psi/M^2}}.\end{aligned} \tag{20}$$

Multiplying both sides of Eq. (20) by $d\Psi/d\xi$, and then integrating it for $\xi$, we can derive the differential equation of $\Psi$ (see Appendix),

$$\frac{1}{2}\left(\frac{d\Psi}{d\xi}\right)^2 + V(\Psi, M) = 0, \tag{21}$$

where $V(\Psi, M)$ is the Sagdeev pseudo-potential [9-13], expressed by

$$\begin{aligned}V(\Psi, M) &= \frac{\eta D_p}{q_p(2q_p-1)(3q_p-2)\gamma} - \frac{\eta \left[1-(q_p-1)\gamma\Psi\right]^{\frac{q_p}{q_p-1}}}{q_p(2q_p-1)(3q_p-2)\gamma}\left\{D_p + \left[2B_p - A_p(3q_p-2)\right]q_p\gamma\Psi + B_p q_p(2q_p-1)\gamma^2\Psi^2\right\} \\ &+ M^2(1-\eta)\left[1 - \left(1 - \frac{2\Psi}{M^2}\right)^{1/2}\right] + \frac{\mu_h D_h(\mu_c + \mu_h\beta)}{q_h(2q_h-1)(3q_h-2)\beta} + \frac{\mu_c D_c(\mu_c + \mu_h\beta)}{q_c(2q_c-1)(3q_c-2)} \\ &- \frac{\mu_h \left[1 + \frac{(q_h-1)\beta\Psi}{\mu_c + \mu_h\beta}\right]^{1/(q_h-1)}}{(3q_h-2)}\left[\frac{D_h(\mu_c + \mu_h\beta)}{q_h(2q_h-1)\beta} + \frac{E_h \Psi}{q_h(2q_h-1)} + \frac{F_h \beta\Psi^2}{(2q_h-1)(\mu_c + \mu_h\beta)} + \frac{B_h(q_h-1)\beta^2\Psi^3}{(\mu_c + \mu_h\beta)^2}\right] \\ &- \frac{\mu_c \left[1 + \frac{(q_c-1)\Psi}{\mu_c + \mu_h\beta}\right]^{1/(q_c-1)}}{(3q_c-2)}\left[\frac{D_c(\mu_c + \mu_h\beta)}{q_c(2q_c-1)} + \frac{E_c \Psi}{q_c(2q_c-1)} + \frac{F_c \Psi^2}{(2q_c-1)(\mu_c + \mu_h\beta)} + \frac{B_c(q_c-1)\Psi^3}{(\mu_c + \mu_h\beta)^2}\right],\end{aligned} \tag{22}$$

where the abbreviations are

$$D_j = 2 + 2B_j - A_j(3q_j - 2) - 7q_j + 6q_j^2, \tag{23}$$

$$E_j = -2 - 2B_j + 9q_j - 13q_j^2 + 6q_j^3 + A_j(3q_j - 2), \tag{24}$$

$$F_j = B_j + A_j(2 - 5q_j + 3q_j^2). \tag{25}$$

Eq. (20) can be treated as an "energy integral" of the oscillating particle with an unit mass and the velocity $d\Psi/d\xi$ at the position $\Psi$ and in the potential $V(\Psi, M)$. According to the Sagdeev pseudo-potential theory, the solitary wave solutions of Eq. (21) exist if the following four conditions [7] are satisfied,

(i) At $\Psi = 0$, $V(\Psi = 0, M) = 0$ and $dV(\Psi, M)/d\Psi|_{\Psi=0} = 0$.

(ii) $\left(d^2 V(\Psi, M)/d\Psi^2\right)_{\Psi=0} < 0$, so that the fixed point at the origin is unstable.

(iii) There is a nonzero $\Psi_m$ at which $V(\Psi_m) = 0$.

(iv) $V(\Psi) < 0$ when $\Psi$ is between 0 and $\Psi_m$.

And further, if $\left(d^3V(\Psi,M)/d\Psi^3\right)_{\Psi=0}>(<)0$, the solitary wave is compressive (rarefactive) [28]. The condition (iii) implies that quasiparticles with zero total energy will be reflected at the position $\Psi = \Psi_m$. The condition (iv) indicates that $V$ must be a potential trough which the quasi-particles can be trapped and experience oscillations.

From the above condition (i) and Eq. (22), we find that only when $q_p \to 1$ and $\alpha_p = 0$ (i.e., the positrons are the Maxwellian distribution), $q_c = q_h$ and $\alpha_c = \alpha_h$ (i.e., the nonextensive parameter and the nonthermal parameter of the cold and hot electrons are the same), the equation has the solitary wave solutions. In this case, we no longer distinguish the subscripts of the nonextensive parameter $q$ and the nonthermal parameter $\alpha$ for the cold and hot electrons, and then the Sagdeev pseudo-potential (22) can be reduced to that,

$$V(\Psi,M) = \frac{\eta}{\gamma}\left[1-\exp(-\gamma\Psi)\right] + M^2(1-\eta)\left[1-\left(1-\frac{2\Psi}{M^2}\right)^{\frac{1}{2}}\right] + \frac{D(\mu_c+\mu_h\beta)(\mu_h\beta^{-1}+\mu_c)}{q(2q-1)(3q-2)}$$
$$-\frac{\left[1+\frac{(q-1)\beta\Psi}{\mu_c+\mu_h\beta}\right]^{1/(q-1)}}{(3q-2)}\left\{\frac{(\mu_h+\mu_c)}{q(2q-1)}\left[D(\mu_c\beta^{-1}+\mu_h)+E\Psi\right]+\frac{F\Psi^2}{2q-1}+B(q-1)\frac{(\mu_h\beta^2+\mu_c)\Psi^3}{(\mu_c+\mu_h\beta)^2}\right\}.$$

(26)

It is obvious that when we take limit of $q_j \to 1$ and $\alpha_j = 0$, (26) becomes the Sagdeev pseudo-potential for the plasma with a Maxwellian distribution [7], namely,

$$V(\Psi,M)\big|_{q\to1,\alpha=0} = \frac{\eta}{\gamma}\left[1-\exp(-\gamma\Psi)\right] + M^2(1-\eta)\left[1-\left(1-\frac{2\Psi}{M^2}\right)^{\frac{1}{2}}\right]$$
$$+\mu_c(\mu_c+\mu_h\beta)\left[1-\exp\left(\frac{\Psi}{\mu_c+\mu_h\beta}\right)\right]+\frac{\mu_h}{\beta}(\mu_c+\mu_h\beta)\left[1-\exp\left(\frac{\beta\Psi}{\mu_c+\mu_h\beta}\right)\right].$$

(27)

When the condition (ii) is applied to Eq. (26), it is easy to find the existence condition for the solitary wave local structure, which requires the Mach number to satisfy

$$M^2 > \frac{q(2q-1)(3q-2)(\eta-1)}{D(q-2)-2E-2qF-q(2q-1)(3q-2)\eta\gamma}.$$

(28)

So we have that

$$M > M_{\min} = \sqrt{\frac{q(2q-1)(3q-2)(\eta-1)}{D(q-2)-2E-2qF-q(2q-1)(3q-2)\eta\gamma}},$$

(29)

where $M_{\min}$ is the minimum limit of $M$, below which there is no the solitary wave. Amplitude of the solitary waves tends to zero as the Mach number $M$ tends to $M_{\min}$. In the case of $\alpha = 0$ and $q \to 1$, the $M_{\min}$ becomes that in the plasma with a Maxwellian distribution [52,53],

$$M_{\min} = \sqrt{\frac{1-\eta}{1+\eta\gamma}}.$$

(30)

The maxmum limit $M_{\max}$ of Mach number $M$ can be found by imposing the condition $V(\Psi_m, M_{\max}) \geq 0$ [7,46,47], where $\Psi_m = M_{\max}^2/2$ is the maximum value that makes the cold ion density real. Therefore, from Eq. (26) we get that

$$V(\Psi_m, M_{\max}) = \frac{\eta}{\gamma}\left[1-\exp\left(-\frac{\gamma M_{\max}^2}{2}\right)\right] + M_{\max}^2(1-\eta) + \frac{D(\mu_c+\mu_h\beta)}{q(2q-1)(3q-2)}\left(\frac{\mu_h}{\beta}+\mu_c\right)$$

$$-\frac{\left[1+\frac{(q-1)\beta M_{max}^2}{2(\mu_c+\mu_h\beta)}\right]^{1/(q-1)}}{(3q-2)}\left\{\frac{(\mu_h+\mu_c)}{q(2q-1)}\left[D(\mu_c\beta^{-1}+\mu_h)+E\frac{M_{max}^2}{2}\right]+\frac{FM_{max}^4}{4(2q-1)}+B(q-1)\frac{(\mu_h\beta^2+\mu_c)}{(\mu_c+\mu_h\beta)^2}\frac{M_{max}^6}{8}\right\}\geq 0. \quad (31)$$

The allowable range of the Mach number for the solitary waves is determined by Eq. (28) and Eq. (31). From Eq. (26), we can further find the condition for the compressive or rarefactive solitary waves, namely,

$$\frac{d^3V(\Psi,M)}{d\Psi^3}\Big|_{\Psi=0}=-\frac{(\mu_c+\mu_h\beta^2)}{(3q-2)(\mu_c+\mu_h\beta)^2}\left[6B(q-1)+\frac{6F}{(2q-1)}+\frac{3E(2-q)}{q(2q-1)}+\frac{D(2-q)(3-2q)}{q(2q-1)}\right]$$
$$+\frac{3(1-\eta)}{M^4}+\eta\gamma^2 \begin{cases} >0, \text{ for compressive waves.} \\ <0, \text{ for rarefactive waves.} \end{cases} \quad (32)$$

## 3. Numerical Analyses and Discussions

From the hydrodynamic equations (1)-(3), we have derived the differential equation (21) for the normalized electrostatic potential $\Psi$ and its Sagdeev pseudo-potential (26) with solitary waves in the four-component plasma with the two-temperature electrons which follow the CT distribution. And based on the Sagdeev pseudo-potential theory, we have derived the existence condition for the solitary wave solutions in Eq. (21). The condition equals to that the Mach number satisfy the inequality, $M_{min} < M < M_{max}$, where $M_{min}$ and $M_{max}$ can be determined by Eq. (29) and Eq. (31) respectively. Further we have found the conditions (32) for the solitary waves to be compressive or rarefactive ones.

In order to see more clearly the properties of the Sagdeev pseudo-potential, the solitary wave solutions, the existence condition for the solitary waves and the condition for the solitary waves to be compressive or rarefactive, now we make the numerical analyses. For this purpose, we first choose some appropriate physical parameters in the plasma, such as $\eta = 0.34$, $\mu_c = 0.1$, $\mu_h = 0.9$, $\gamma = 0.04$, $\beta = 0.05$ and $M = 0.82$.

In Fig.1, (a)-(c), we show the Sagdeev pseudo-potential $V(\Psi, M)$ in Eq. (26) as a function of the normalized electrostatic potential $\Psi$ for different nonextensive parameter $q$ and the nonthermal parameter $\alpha$. Fig.1(a) is $V(\Psi, M = 0.82)$ as a function of $\Psi$ for a fixed $\alpha = 0.25$ and four different $q$, which show that with the increase of negative $\Psi$ (the absolute value of $\Psi$ decreases), $V(\Psi)$ will decrease monotonously. As the positive $\Psi$ increases, $V(\Psi)$ shows its maximum value and then gradually decreases. When $q < 1$, increasing $q$ will cause the more obvious change of $V(\Psi)$ with $\Psi$. But when $q > 1$, increasing $q$ will make the change of $V(\Psi)$ with $\Psi$ less obvious.

Fig.1(b) and (c) are $V(\Psi, M = 0.82)$ as a function of $\Psi$ for a fixed $q$ and three different $\alpha$, where (b) is for the case of $3/5 < q < 1$ and (c) is for the case of $q > 1$, showing the significant differences between the cases of $q > 1$ and $3/5 < q < 1$. It is shown that with the increase of $\alpha$, $V(\Psi)$ increases basically for $3/5 < q < 1$, but it firstly increases and then decreases for $q > 1$. And $V(\Psi)$ as a function of $\Psi$ is significantly different from the case in the plasma with a Maxwellian distribution.

If we give the initial condition as $\Psi(\xi = 0) = 0$, the stationary normalized electrostatic potential $\Psi(\xi)$ can be calculated by making numerical integration for Eq. (21). The numerical results are shown in Figs. 2(a)-(b). It is clear that the potential $\Psi(\xi)$ depends significantly on the parameters $q$ and $\alpha$, and so it is different from that in the plasma with a Maxwellian distribution.

Fig. 2(a) is $\Psi(\xi)$ as a function of $\xi$ for a fixed $\alpha = 0.06$ and four different $q$ in the plasma,

where two values of *q* are taken less than 1 and the other two values of *q* are taken greater than 1. It reveals that the present plasma supports both compressive ($\Psi > 0$) and rarefactive ($\Psi < 0$) solitary waves. For the case of $3/5 < q < 1$, the amplitude of the solitary waves increases with the increase of *q*. But for the case of $q > 1$, the amplitude of the solitary waves decreases with the increase of *q*.

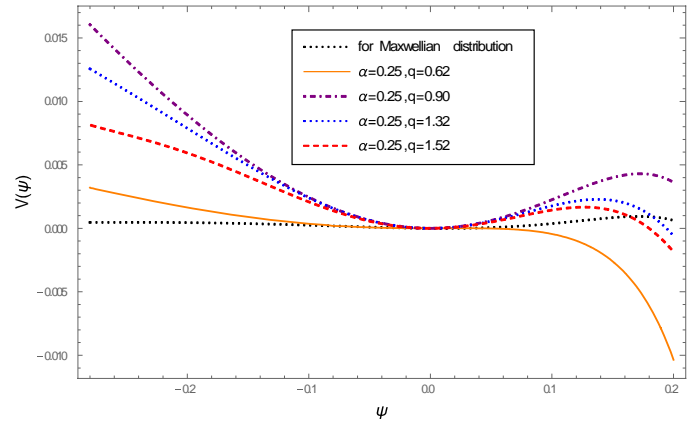

(a)

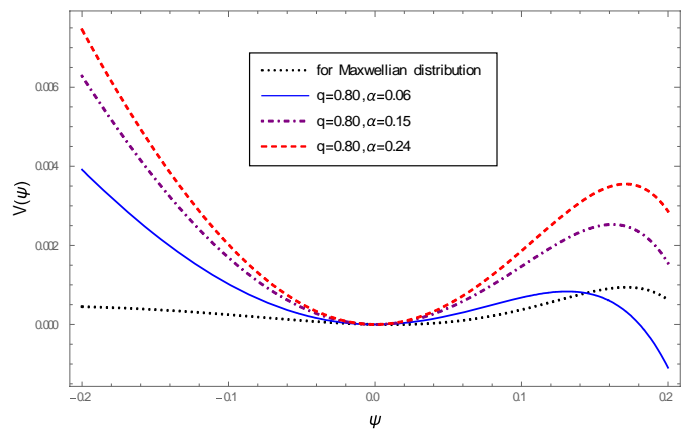

(b)

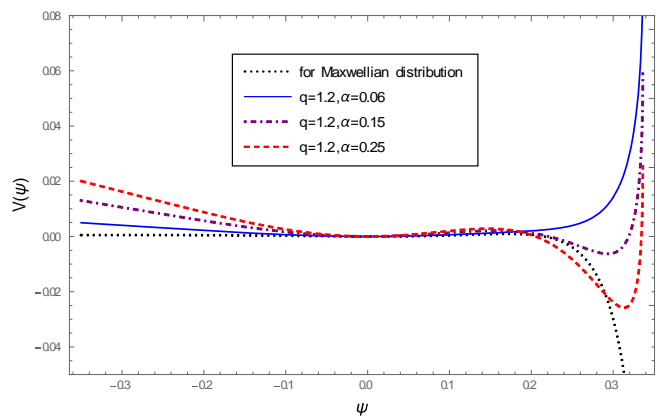

(c)

**Fig. 1.** Dependence of Sagdeev pseudo-potential $V(\Psi)$ on the parameters *q* and *α*.

Fig. 2(b) is $\Psi(\xi)$ as a function of $\xi$ for a fixed *q* and two different values of *α* in the plasma,

where the fixed $q = 0.81$ is taken for the case of $3/5 < q < 1$ and the fixed $q = 1.205$ is taken for the case of $q > 1$. It is shown that for the case of $q = 0.81$, both the compressive ($\Psi > 0$) and rarefactive ($\Psi < 0$) solitary wave coexist in the present plasma, and the amplitude of the waves decreases with the increase of $\alpha$, but for the case of $q = 1.205$, only the rarefactive solitary wave ($\Psi < 0$) exists in the plasma and the amplitude of the waves also decreases with the increase of $\alpha$.

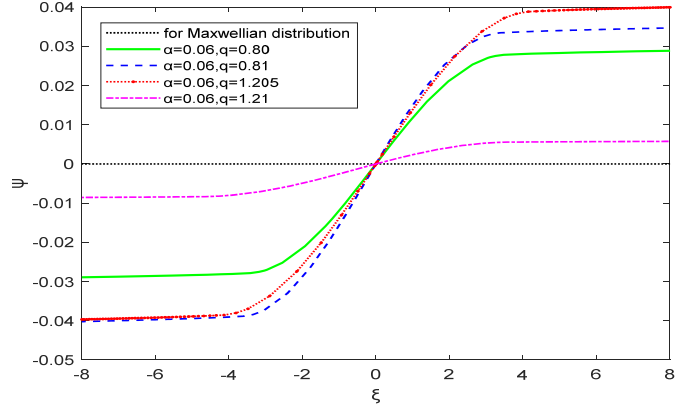

**Fig. 2(a).** Dependence of $\Psi(\xi)$ as a function of $\xi$ on parameters $q$ for a fixed $\alpha$.

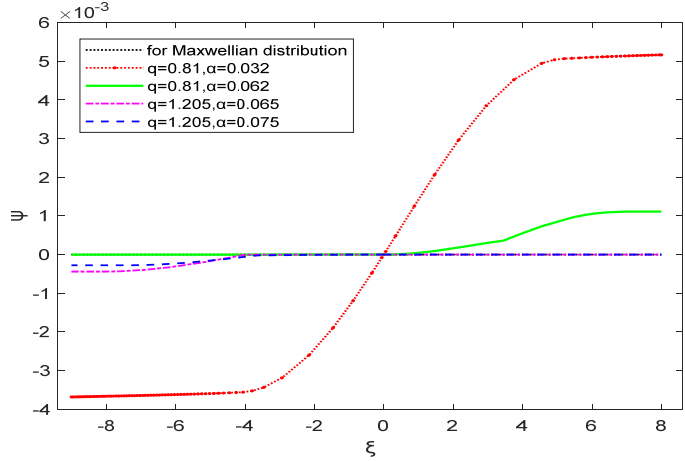

**Fig. 2(b).** Dependence of $\Psi(\xi)$ as a function of $\xi$ on the parameter $\alpha$ for a fixed $q$.

The existence condition for the solitary wave solutions in Eq. (21) equals to that the Mach number satisfy $M_{\min} < M < M_{\max}$, where the minimum Mach number $M_{\min}$ and the maximum Mach number $M_{\max}$ is determined by Eq. (29) and Eq. (31) respectively. Based on Eq. (29), we can analyze numerically the dependence of the minimum Mach number $M_{\min}$ on the nonextensive parameter $q$ and the nonthermal parameter $\alpha$ in the plasma. In Fig. 3, we give $M_{\min}$ as a function of $q$ for three different $\alpha$. It is shown that when $q$ is small, $M_{\min}$ increases rapidly to reach a peak with the increase of $q$, and then with the increase of $q$, $M_{\min}$ decreases gradually. It is also shown that when $q$ is small (viz, $q < 0.81$), $M_{\min}$ hardly varies with the increase of $\alpha$, but when $q$ is large (viz, $q > 0.81$), $M_{\min}$ increases rapidly with the increase of $\alpha$.

Based on Eq. (31), we can analyze numerically the dependence of the Sagdeev pseudo-potential $V(\Psi_m, M_{max})$ on the maximum Mach number $M_{max}$ for certain nonextensive parameter $q$

> 3/5 and certain nonthermal parameter $\alpha > 0$ in the plasma.

Fig.4 is $V(\Psi_m, M_{max})$ as a function of $M_{max}^2/2$ for $\alpha = 0.35$, $q = 0.8$ and $q = 1.2$ respectively. It is shown that $V(\Psi_m, M_{max})$ is always negative for any $M_{max} = 0 \sim \infty$, so there is no any restriction on $M_{max}$ for the existence of the solitary waves in the present plasma.

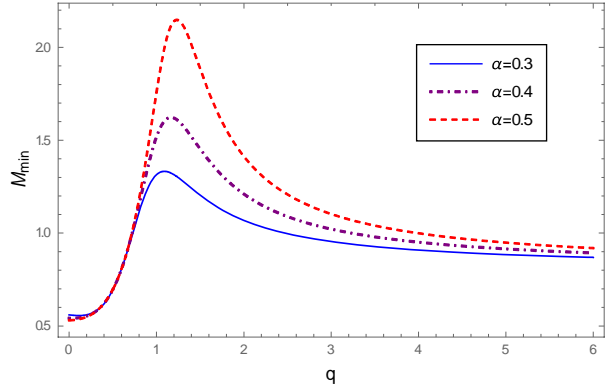

**Fig. 3.** $M_{min}$ as a function of $q$ for three different $\alpha$.

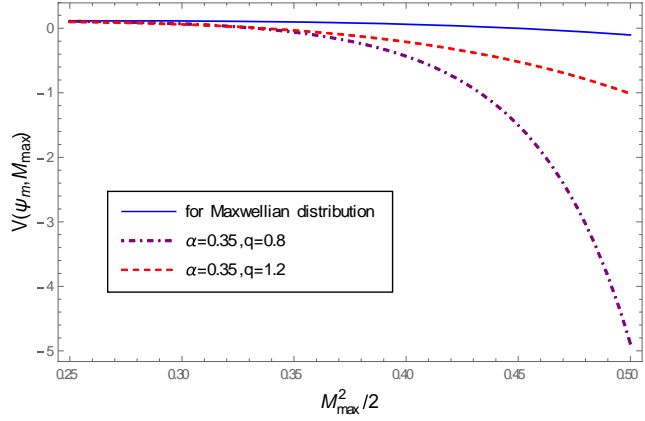

**Fig. 4.** $V(\Psi_m, M_{max})$ as a function of $M_{max}^2/2$ for certain $\alpha$ and $q$.

We have numerically analyzed $(d^3V/\partial\Psi^3)_{\Psi=0}$ in Eq. (32) so as to show whether the solitary waves are compressive or rarefactive for different nonextensive parameter $q$ and nonthermal parameter $\alpha$. Fig. 5 is $(d^3V/\partial\Psi^3)_{\Psi=0}$ based on Eq. (32) as a function of $q$ four three different values of $\alpha$. It is shown that for $3/5 < q < 1$, $(d^3V/\partial\Psi^3)_{\Psi=0} < 0$, so there are only rarefactive solitary waves in the present plasma, and it is basically independent of $\alpha$, but for $q > 1$, $(d^3V/\partial\Psi^3)_{\Psi=0}$ can be either greater than zero or less than zero, so there can be both compressive and rarefactive solitary waves.

Furthermore, we observe that the critical value of the transition from rarefactive to compressive solitary waves significantly depends on $q$ and $\alpha$. As $q$ and $\alpha$ increase, the critical value gradually increases. This indicates that when the values of $q$ and $\alpha$ are less than the critical value, there are only rarefactive solitary waves in the system, and when the value is greater than the critical value, there are only compressive solitary waves.

The black dotted line in Fig. 5 represents the case when the nonthermal property of the electrons vanishes (i.e., $\alpha = 0$). It shows that the present system supports the existence of rarefactive solitary waves for $3/5 < q < 1$, and for $q > 1$, it almost only supports the existence of compressive solitary waves. If we take the limits $q \to 1$ and $\alpha = 0$, it will return to Maxwellian solitary wave. As shown in Fig. 5, it is a rarefactive wave.

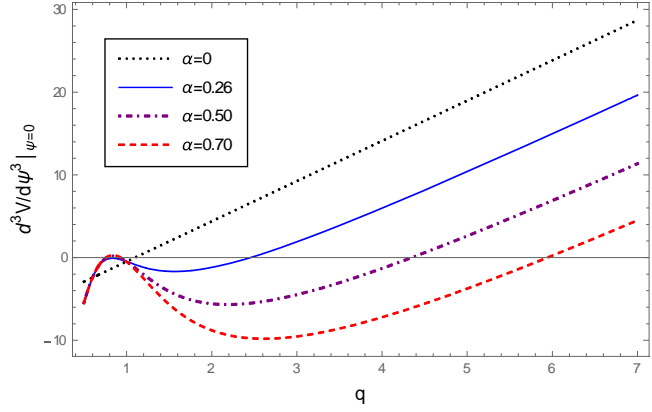

**Fig.5** The condition for compressive or rarefactive solitary waves as a function of $q$ for three different $\alpha$.

## 4. Small amplitude analysis

The properties of large amplitude solitons have been presented above. In this section, we discuss more characteristics of solitons with the small amplitude limit $\Psi \ll 1$. It is useful to expand the Sagdeev potential represented by Eq. (26). Retaining only to the third order in $\Psi$, we get

$$V(\Psi) \simeq P_1\Psi + P_2\Psi^2 + P_3\Psi^3 + o(\Psi^4), \qquad (33)$$

where

$$P_1 = 0, \quad P_2 = \frac{-3 - 36\alpha + q(14 - 15q + 40\alpha)}{2(3 - 14q + 15q^2 + 12\alpha)} - \frac{\eta - 1 + M^2\eta\gamma}{2M^2}, \qquad (34)$$

$$P_3 = -\frac{\left[-3E(q-2) + D(q-2)(2q-3) + 6q(B + F - 3Bq + 2Bq^2)\right](\mu_c + \mu_h\beta^2)}{6q(2 - 7q + 6q^2)(\mu_c + \mu_h\beta)^2} + \frac{1-\eta}{2M^4} + \frac{1}{6}\eta\gamma^2. \qquad (35)$$

Applying the soliton solution in Eq. (33), Eq. (21) is evolved into:

$$\frac{1}{2}\left(\frac{d\Psi}{d\xi}\right)^2 = -\Psi^2(P_2 + P_3\Psi). \qquad (36)$$

It can be seen that $\Psi$ has real function requiring $P_2 + P_3\Psi < 0$. Assuming $P_2$ is positive, Eq. (36) is integrated and a solitary wave solution is obtained:

$$\Psi(\xi) = -\frac{P_2}{P_3}\text{Sech}^2\left(\frac{\xi}{\sqrt{2/P_2}}\right). \qquad (37)$$

From the solitary solution (37), we find that the amplitude and width of the solitary waves are $\Psi_m = -P_2/P_3$ and $\Delta = (2/P_2)^{1/2}$, respectively. Therefore, the solitary wave is compressive or rarefactive depends on the signs of $P_2$ and $P_3$. On the one hand, if $P_2 > 0$, the type of solitary waves depends only on $P_3$. When $P_3 > 0$, it is a rarefactive solitary wave, otherwise it is a compressive wave, and the width of the solitary waves is $\Delta = (2/P_2)^{1/2}$. On the other hand, if $P_2 < 0$, the width $\Delta = (-2/P_2)^{1/2}$ is valid. At this time, if $P_3 > 0$, it is a compressive solitary wave, otherwise it is a rarefactive solitary wave [7].

## 5. Summary and Conclusions

In summary, we have studied the ion acoustic solitary waves in a general four-component plasma consisting of the cold fluid ions, positrons, cold electrons and hot electrons (the two-temperature electrons), where the two-temperature electrons and positrons both follow the Cairns-Tsallis distribution and have different nonextensive and nonthermal parameters.

Based on the continuity equation (1), the equation (2) of fluid motion and the Poisson equation (3) in the plasma, we have derived differential equation (21) for the normalized electrostatic potential $\Psi$ and its related Sagdeev pseudo-potential (22). According to the existence conditions of solitary waves, we have found that solitary waves exist only when the positrons reduce to the Maxwellian distribution, i.e. $q_p \rightarrow 1$, $\alpha_p = 0$, and the nonextensive parameter $q$ and the nonthermal parameter $\alpha$ of cold electrons and hot electrons are the same, i.e. $q_c = q_h$, $\alpha_c = \alpha_h$.

And based on the Sagdeev pseudo-potential (26) with solitary waves, we have further derived the condition for the solitary wave solutions to exist in Eq. (21). The condition is equivalent to a restriction on the Mach number $M$, i.e. the inequality, $M_{min} < M < M_{max}$, where the maximum Mach number $M_{min}$ and the minimum Mach number $M_{max}$ depend strongly on the nonextensive parameter $q$ and nonthermal parameter $\alpha$, and they can be determined by Eq. (29) and Eq. (31) respectively. Further we have found the condition (32) for the solitary waves to be compressive or rarefactive ones.

In order to study the ion acoustic solitary waves in the plasma more clearly, the numerical analyses of the above quantities have been made. The numerical results are given by Figs.1(a)-(c), Figs.2(a)-(b), Fig.3, Fig.4 and Fig.5, respectively. From these figures we have shown that all the properties of ion acoustic solitary waves are significantly dependent on the nonextensive parameter $q$ and nonthermal parameter $\alpha$ of the Cairns-Tsallis distribution in the plasma, and therefore they are generally different from those in the same plasma following a Maxwellian distribution. In addition, we find that there is no any restriction on $M_{max}$ for the existence of the solitary waves in the present plasma.

Finally, we discussed a small amplitude ion acoustic wave through the Sagdeev pseudopotential analysis. With the small amplitude limit $\Psi \ll 1$, the normalized electrostatic potential of ion acoustic waves is theoretically solved and given in Eq. (39). It further clears that our plasma model can admit compressive as well as rarefactive ion acoustic solitons.

This research has great scientific research and application value in revealing the law of wave propagation, accurately revealing natural phenomena and determining the properties of physical materials.

**Acknowledgements**

This work is supported by the National Natural Science Foundation of China under Grant No. 11775156.**Appendix**

The derivation of Sagdeev pseudo-potential (22):

Multiplying both sides of Eq. (20) by d$\Psi$/d$\xi$ and integrating it, we have that the left side of (22) is

$$\int \frac{d\Psi}{d\xi} \frac{d^2\Psi}{d\xi^2} d\xi = \frac{1}{2}\left(\frac{d\Psi}{d\xi}\right)^2, \tag{A.1}$$

and the right side of (22) is

$$\int \left[ n_h + n_c - \eta n_p - (1-\eta) n_i \right] \frac{d\Psi}{d\xi} d\xi$$

$$= -\eta \int \left[ 1 - (q_p - 1)\gamma\Psi \right]^{\frac{1}{q_p - 1}} \left[ 1 - A_p \gamma\Psi + B_p (\gamma\Psi)^2 \right] d\Psi - (1-\eta) \int \frac{d\Psi}{\sqrt{1 - 2\Psi/M^2}}$$

$$+ \int \mu_h \left[ 1 + \frac{(q_h - 1)\beta\Psi}{\mu_c + \mu_h \beta} \right]^{\frac{1}{q_h - 1}} \left[ 1 + \frac{A_h \beta\Psi}{\mu_c + \mu_h \beta} + B_h \left( \frac{\beta\Psi}{\mu_c + \mu_h \beta} \right)^2 \right] d\Psi \quad\quad (A.2)$$

$$+ \int \mu_c \left[ 1 + \frac{(q_c - 1)\Psi}{\mu_c + \mu_h \beta} \right]^{\frac{1}{q_c - 1}} \left[ 1 + \frac{A_c \Psi}{\mu_c + \mu_h \beta} + B_c \left( \frac{\Psi}{\mu_c + \mu_h \beta} \right)^2 \right] d\Psi.$$

On the right side of Eq. (A.2), the four integrals are calculated, respectively, as

$$-\eta \int \left[ 1 - (q_p - 1)\gamma\Psi \right]^{\frac{1}{q_p - 1}} \left[ 1 - A_p \gamma\Psi + B_p (\gamma\Psi)^2 \right] d\Psi = -\frac{\eta D_p}{q_p (2q_p - 1)(3q_p - 2)\gamma}$$

$$+ \frac{\eta \left[ 1 - (q_p - 1)\gamma\Psi \right]^{\frac{q_p}{q_p - 1}}}{q_p (2q_p - 1)(3q_p - 2)\gamma} \left\{ D_p + \left[ 2B_p - A_p (3q_p - 2) \right] q_p \gamma\Psi + B_p q_p \left[ 2q_p - 1 \right] \gamma^2 \Psi^2 \right\}, \quad (A.3)$$

$$-\int (1-\eta) \left( 1 - \frac{2\Psi}{M^2} \right)^{-\frac{1}{2}} d\Psi = M^2 (1-\eta) \left[ \left( 1 - \frac{2\Psi}{M^2} \right)^{\frac{1}{2}} - 1 \right]; \quad\quad (A.4)$$

$$\int \mu_h \left[ 1 + \frac{(q_h - 1)\beta\Psi}{\mu_c + \mu_h \beta} \right]^{1/(q_h - 1)} \left[ 1 + \frac{A_h \beta\Psi}{\mu_c + \mu_h \beta} + B_h \left( \frac{\beta\Psi}{\mu_c + \mu_h \beta} \right)^2 \right] d\Psi = -\frac{\mu_h D_h (\mu_c + \mu_h \beta)}{q_h (2q_h - 1)(3q_h - 2)\beta}$$

$$+ \frac{\mu_h \left[ 1 + \frac{(q_h - 1)\beta\Psi}{\mu_c + \mu_h \beta} \right]^{1/(q_h - 1)}}{(3q_h - 2)} \left[ \frac{D_h (\mu_c + \mu_h \beta)}{q_h (2q_h - 1)\beta} + \frac{E_h \Psi}{q_h (2q_h - 1)} + \frac{F_h \beta\Psi^2}{(2q_h - 1)(\mu_c + \mu_h \beta)} + \frac{B_h (q_h - 1)\beta^2 \Psi^3}{(\mu_c + \mu_h \beta)^2} \right], \quad (A.5)$$

$$\int \mu_c \left[ 1 + \frac{(q_c - 1)\Psi}{\mu_c + \mu_h \beta} \right]^{1/(q_c - 1)} \left[ 1 + \frac{A_c \Psi}{\mu_c + \mu_h \beta} + B_h \left( \frac{\Psi}{\mu_c + \mu_h \beta} \right)^2 \right] d\Psi = -\frac{\mu_c D_c (\mu_c + \mu_h \beta)}{q_c (2q_c - 1)(3q_c - 2)}$$

$$+ \frac{\mu_c}{(3q_c - 2)} \left[ \frac{D_c (\mu_c + \mu_h \beta)}{q_c (2q_c - 1)} + \frac{E_c \Psi}{q_c (2q_c - 1)} + \frac{F_c \Psi^2}{(2q_c - 1)(\mu_c + \mu_h \beta)} + \frac{B_c (q_c - 1)\Psi^3}{(\mu_c + \mu_h \beta)^2} \right] \left[ 1 + \frac{(q_c - 1)\Psi}{\mu_c + \mu_h \beta} \right]^{1/(q_c - 1)}, \quad (A.6)$$

where $D_j = 2 + 2B_j - A_j (3q_j - 2) - 7q_j + 6q_j^2$, $E_j = -2 - 2B_j + 9q_j - 13q_j^2 + 6q_j^3 + A_j (3q_j - 2)$, and $F_j = B_j + A_j (2 - 5q_j + 3q_j^2)$.

Substituting Eqs. (A.3)-(A.6) into Eq. (A.2), we obtain Eq. (21).